\documentclass{svproc}
\usepackage[latin1]{inputenc}
\usepackage[T1]{fontenc}
\usepackage{amsmath}
\usepackage{amsfonts}
\usepackage{amssymb}
\usepackage{graphicx}
\usepackage[pagebackref=true,hyperindex=true]{hyperref}

\usepackage{xcolor}

\usepackage{draftwatermark}
\SetWatermarkText{DRAFT}
\SetWatermarkScale{3}
\SetWatermarkLightness{0.9}

\begin{document}
	
	\title{Impact of the Interaction Network on the Dynamics of 
		Word-of-Mouth with Information Seeking}
	
	\titlerunning{Dynamics of Word-of-Mouth with Information Seeking}
	
	\author{Samuel Thiriot}
	\institute{EIFER, European Institute for Energy Research\\
		Emmy-Noether-Straße 11,
		76131 Karlsruhe,
		Germany\\
		\email{samuel.thiriot@res-ear.ch}
	}
	\maketitle
	
	\bibliographystyle{plain}
	
	\begin{abstract}
	Word-of-Mouth refers to the dynamics of interpersonal communication occurring during the diffusion of innovations (novel practices, ideas or products). 
	According to field studies, word-of-mouth is made of both information seeking and proactive communication: individuals first become aware of the existence of an innovation, then start actively seeking out for the expert knowledge required to evaluate the innovation; when they hold the expert knowledge, they might start promoting it pro-actively. Successful diffusion of innovation requires the individuals to hold both awareness and expert knowledge, so they can evaluate the innovation and use it properly.
	A computational model "USA/IPK" was recently proposed to study the role and impact of information seeking on the dynamics of word-of-mouth.
	We propose here an analysis of the impact of the network of interaction on the dynamics of this model.
	We compare the dynamics of the model over networks generated with different algorithms with the original dynamics. 
	The results demonstrate the dynamics of the model are similar across tested networks, with the noticeable exception of the efficiency of the diffusion which varies between networks having similar densities and sizes.
	\end{abstract}
	
	\section{Introduction}
	
	\subsection{Word-of-mouth: Evidence on Information Seeking}
	
	When individuals discuss an innovation (a novel product, practice, idea) \cite{samuel_thiriot:bib_customer_value:rogers_2003_1}, they \textit{spread the word} about its existence and qualities. More people become aware of a product through word-of-mouth than traditional advertisement \cite{Sheth1971}. Most consumers attribute a higher importance to interpersonal influence than other sources \cite{Katz1955}.
	As a consequence, word-of-mouth is consistently said to determine the success or failure of innovations~\cite{samuel_thiriot:bib_customer_value:rogers_2003_1} and facilitate the diffusion of products~\cite{Katz1955}.
	
	Word-of-mouth was often reduced to an epidemic process in which individuals ``contaminate'' each other with information \cite{Goffman1964}.	
	Yet interpersonal communication about innovations or products \cite{samuel_thiriot:bib_customer_value:arndt_1967_1} does not only include the proactive emission of information, but also communications initiated by people who seek out information about an innovation \cite{Gilly1998,samuel_thiriot:bib_customer_value:rogers_2003_1}. 
	When an individual discovers the existence of the innovation, that is when he receives awareness knowledge from advertisement or another individual, he might engage (or not) in information seeking depending on his characteristics \cite{Gilly1998,samuel_thiriot:bib_customer_value:rogers_2003_1}.
	Expert knowledge covers ``how to'' use the innovation (know-how knowledge) \cite{samuel_thiriot:bib_customer_value:rogers_2003_1}, ``why'' the innovation works (principles-knowledge) \cite{samuel_thiriot:bib_customer_value:rogers_2003_1}, product category 
or product-class knowledge 
or brand knowledge 
.
	This expert knowledge
might be gathered from individuals who hold it prior to the diffusion of the innovation because of their education or training, because they read specialized press, had experience with another product of same brand, category or class, or because they received this information from another individual.
	Once they hold the expert knowledge, people might engage into pro-actively passing the word around about the innovation, for instance because they are willing to help others \cite{Gilly1998} or because they are satisfied or dissatisfied after adoption \cite{samuel_thiriot:bib_customer_value:anderson_1998_1}.
	
	Information seeking stands as a step required for most individuals to be able to decide to adopt or reject a product \cite{Katz1955}.   
	In the case of the diffusion of disruptive innovations such as vaccination or contraceptives \cite{samuel_thiriot:bib_customer_value:rogers_2003_1}, information seeking is even seen as a mandatory step for individuals to adopt the innovation, as it enables people to understand why it works and how to use it. For instance, parents do not accept the vaccination of their children without gathering more knowledge first \cite{Trim2012}.
	Even if innovations might be adopted without expert knowledge, the misuse of the innovation may later cause its discontinuance \cite{samuel_thiriot:bib_customer_value:rogers_2003_1}. 
	As a consequence, \textit{a company or organization promoting an innovation attempts to maximize the proportion of the population which is not only aware, but also holds expertise on the innovation} \cite{samuel_thiriot:bib_customer_value:rogers_2003_1}.
	
	\subsection{Word-of-mouth: Computational Models}
		
	Numerous computational or mathematical models \cite{samuel_thiriot:bib_customer_value:meade_2006_1,Peres2010,Kiesling2012} were designed to understand the diffusion of information and innovations \cite{Granovetter1978,samuel_thiriot:bib_customer_value:goldenberg_2000_1,samuel_thiriot:bib_customer_value:leskovec_2007_1}, assess the potential diffusion of products~\cite{Chatterjee1990}, and recommend strategies to accelerate or maximize this diffusion \cite{Kempe2003}. The three main types of models related to information diffusion are based on information cascades, social influence and social learning \cite{samuel_thiriot:bib_customer_value:leskovec_2007_1,Young2009}. 
	The models developed in these last two categories describe the flow of influence within the population, without explicitly representing the flow of information, and can not be used to study the impact of information seeking.
%
%
	Marketing models based on \textit{information cascades} \cite{Goffman1964,samuel_thiriot:bib_sma_simulation:goldenberg_2001_1,Kempe2003}
	 rely on an analogy with epidemic models \cite{Goffman1964,samuel_thiriot:bib_sma_simulation:daley_1964_1} such as the SIR model  \cite{samuel_thiriot:bib_sma_simulation:kermack_1927_1,samuel_thiriot:bib_sma_simulation:bailey_1957_1}:
	every individual is either in state Susceptible (no information), Infective (informed and pro-actively passing the information to others) or Recovered (informed but passive). A Susceptible individual becomes Infective ($\text{S}\rightarrow\text{I}$) when he meets an Infective individual. After a given time, Infective agents become Recovered ($\text{I}\rightarrow\text{R}$).
	When enough individuals are passing the word around, information cascades appear in the simulations, as observed in reality \cite{samuel_thiriot:bib_customer_value:leskovec_2007_1}. 
	In this case, the cumulated curve of the proportion of people informed in time follows the traditional S-shaped curve. 
	Unfortunately, because they only include the information passing behaviour without any information seeking, these models only capture part of the dynamics of word-of-mouth.

	A computational model was recently proposed to explore the dynamics of word-of-mouth with information seeking \cite{Thiriot2018}. 
	In this model built as an evolution of the SIR model, two dimensions of knowledge are distinguished: \textit{awareness knowledge} which refers to knowing the innovation exists, and \textit{expert knowledge} which allows the actual understanding of the innovation. 
	The agents representing the individuals are associated with states of information for awareness and expertise, and are associated with traits determining their behaviour regarding information (e.g. curious agents will seek out information when they become aware). 
	In a typical simulation, the population is initialized as unaware of the innovation. A given proportion of agents $k$ is initialized with the expert knowledge, which they are supposed to know because of their education, training or experience with another similar innovation. A low proportion of expertise $k$ in the population is said to represent a disruptive innovation; a higher proportion of expertise represents an incremental innovation that most people can understand as soon as it is advertised. 
	During N steps of the simulation, an advertisement campaign dispatches the awareness knowledge to a small proportion of the population. 
	Then the interpersonal interactions in the model drive the propagation of knowledge. Agents who became aware of the innovation, and have the trait curious, start to seek out for expert knowledge around them; by doing so, they propagate awareness around them; when they discover the expert knowledge, they stop seeking out information; the individuals having trait enthusiastic start promoting the innovation, whilst the others become passive.  
	A typical simulation exhibits two S-shaped curves:
	the higher is the curve of awareness. The lower one is the curve of people who hold both awareness and expert knowledge. 
	This model enables to experiment on the dynamics of information seeking: what are the relative impacts of curious people and enthusiastic ones? On which conditions can we enhance the retrieval of expertise in the population? 
	Published results highlight a surprisingly high efficiency of information seeking, which enables most of the population to achieve gathering an expert knowledge held by as few as 0.1\% of the population. Simulations experiments also suggest three different regimes for different proportions of initial knowledge, suggesting different communication strategies for disruptive and incremental innovations. 
	
\subsection{Research question}

	Each agent in a simulation of the USA/IPK model is located over an interaction network which determines with which agents it interacts at each time step. 
	The dynamics of the model published so far \cite{Thiriot2018} were only simulated over Watts-Strogatz networks.
	Yet existing knowledge on real social networks suggests they might have different characteristics than WS ones, including a skewed distribution of degrees, a strong modularity or a core-periphery structure. 
	As the structure of interactions usually plays a central role on the dynamics of a model, we question here the impact of the networks of interactions on the dynamics of the IPK/USA model: do different networks lead to different qualitative dynamics? 
	In the next section~\ref{exp_protocol}, we describe an experimental protocol to tackle these questions.
	After a detailed description of the USA/IPK model, we select several network generators reproducing statistical properties observed in real networks. 
	We also detail the implementation of these experiments to facilitate replication.
	In the Results section~\ref{xp_results} p.~\pageref{xp_results}, we depict the results of these computations, and compare the qualitative dynamics of the model in its space of parameters with the published ones. 
	We then discuss (\ref{discussion} p.\pageref{discussion}) the implications of these findings on the methodology to compare computational models over networks, the practical findings for the USA/IPK model, and open novel questions arising from these findings.

\section{Experimental protocol\label{exp_protocol}}

	\subsection{Selected Random Network Generators}
	
	As for an epidemic model and other models in which the dynamics depends on cascades, some statistical properties would obviously impact the dynamics of the model. If the density of the network increase, each agent seeking out expert knowledge is more likely to find it out, and agents pro actively transmitting information would have more impact. In the same way, networks having different sizes, diameters or average path lengths might bias the results because the increased steps required for information to flow in the entire population. 
	In order to avoid these trivial effects of networks on the dynamics, we select network generators which are all compliant with the small-world effect (low density, short average path length) and all have a high clustering (or transitivity rate). We then set the parameters of these random network generators so that they generate networks having close statistical properties for density, average path length, count of vertices and diameter. 
	We retain for these experiments three network generators that are able to reach similar statistical properties. 
	
	\textbf{[WS]} The famous Watts and Strogatz $\beta$-model \cite{samuel_thiriot:bib_sma_simulation:watts_1998_1} requires as parameters $N$ the size of the network, $nei$ the neighbourhood of the original lattice and $p^{rewire}$ the rewiring probability. This algorithm starts with a regular lattice of $N$ nodes in which nodes are connected with their $nei$ neighbours (thus having $2.nei$ degree). It then rewires each link with probability $p^{rewire}$. 
	We define here $n=1000$, $nei=5$, $p=0.055$; this leads to networks having a density of $0.01$, average path length $4.4\pm0.3$ and a clustering rate of $0.47\pm0.02$.

	\textbf{[FF]}
	The Forest Fire model was proposed by Leskovec \cite{samuel_thiriot:bib_sma_simulation:leskovec_2007_1} as an algorithm that creates networks having most of the properties observed in real networks, including communities, skewed distribution of degrees, a core-periphery structure. 
	The network is grown step by step, each new node $A$ being attached to $m$ old nodes. Moreover, each time a new link is created between $A$ and $B$, $A$ explores the outgoing and incoming neighbours of $B$. $A$ create links with outgoing nodes of $B$ with a \textit{forward probability} $p$, and also creates links with incoming nodes of $B$ with probability $p.r$, with $r$ the \textit{backward burning ratio}. As this step is ran recursively, $A$ is said to ``burn'' all the possible links. 
	Here FF is set up with $n=1000$, $fw.prob=0.37$, $bw.factor=0.9$, $ambs=1$. 
	The resulting networks have a density of $0.01\pm0.005$, an average path length of $4.13\pm0.2$, and a clustering rate of $0.26\pm0.2$.
	
	\textbf{[SII]} We also test a simple model\footnote{This algorithm is highly similar to the one proposed by Newman and Girvan \cite{samuel_thiriot:bib_sma_simulation:girvan_2002_1} as a test bed for community detection, in which two parameters drive the probability of existence of links respectively intra and inter communities. Our model facilitates the present experiments due to the guaranteed connectedness of the network.} that creates networks composed of several communities (sets of nodes having a strong density). This model, later named SII for Simple Interconnected Islands, starts by creating $n$ islands of identical size $size$. Each island is a random graph in which links exist with probability $p.in$. Each island is connected with all the other islands with $n.inter$ links, each being created between two nodes randomly picked from each island. Density and transitivity in SSI networks can easily be tuned by varying the $n$, $size$ and $p.in$ parameters, while the average path length may be tuned with $n.inter$. Its distribution of degree is nearly a Poisson-like one (as each island is a random network). This average path length remains low, because all the islands are interconnected. 
	For this study we use $n=24$ islands, $size=42$ nodes per island, $p.in=0.235$ of wiring probability within islands, $n.inter=1$ link between each pair of island. 
	The resulting networks count $1008$ vertices, have a density of $0.0101\pm0.002$, an average path length of $4.36\pm0.02$, and a clustering rate of $0.21\pm0.1$. 
	
	\subsection{Space of parameters for the USA/IPK model}

	In order to introduce information seeking, the authors of the USA/IPK distinguish \textit{awareness knowledge} which refers to the existence of the innovation of interest, and \textit{expert knowledge} which refers to the information required to understand, or use in the right way, the innovation. 
	%
	On the \textit{awareness dimension}, the agent is either Unaware (knows the innovation exists), Seeking (discovered the innovation exists and seeks out for expert knowledge) or Aware (knows the innovation exists but does not actively search for information).
	On the \textit{expertise dimension}, the agent is either Ignorant (does not holds the expert knowledge), Proactive (holds the expert knowlege and shares it around him) or Knowledgeable (passively holds the expert knowledge, without sharing it pro-actively). 

	The authors introduced personality variables which are randomly initialized at the beginning of the population based on ratios part provided as model parameters and constant during the simulation. "\textit{Curious}" individuals refer to agents which, when they receive awareness, shift to the "Seeking" state; non curious agents would transition to "aware" instead. "\textit{Enthusiastic}" agents refer to agents which, when they received the expert knowledge, promote it pro actively around them; non enthusiastic agents would transition directly to Knowledgeable instead. "\textit{Supporter}" agents are those who, when they discover awareness when they are already holding expert knowledge, start promoting the information. 
	%
	We redirect the interested reader to the complete description of the model \cite{Thiriot2018}.

	For each network generator, we explore the same space of parameters of the model as in the original study to facilitate comparisons: the proportion of curious and enthusiastic people is explored in $[0:1]$ by steps of $0.05$. The proportions of supporters include $0.0$, $0.1$ and $0.5$. The proportions of initial knowledge $k$ include $k=0.01$ (1\%, disruptive innovation), $k=0.1$ (incremental innovation), $k=0.5$ (standard product). 
	For each point of the space of parameters, $10$ simulations are started with a different random seed. We measure the final proportion of the population holding both the awareness and the expert knowledge. 
	
	The interest of the dynamics of the initial study can not be reduced to one unique epidemic threshold; the interesting elements were mostly qualitative, as the observation of three different regimes depending on the proportion of initial expertise, or the asymmetry in the role of information seeking and proactive communication. As a consequence, we drive an extensive exploration of the space of parameters and interpret the dynamics of the model in this space as in the original study. We will come back on every qualitative finding of the original study and compare the result with different networks.
	
	\subsection{Implementation}
	These experiments require the chaining of the generation of complex networks and the agent-based simulation over this network. For each simulation experiment, a different random seed is defined, a network is generated, its statistical properties are analyzed, then the network is loaded by the simulation model and the simulation is ran until the end of the diffusion dynamics. 
	In order to facilitate the exploration of this experimental design, facilitate reproducibility and avoid manipulation errors, we use the OpenMole scientific software \cite{reuillon2013openmole} which enables the description of computation workflows in which tasks are chained with each other. 
	The generation of the networks is done in R using the igraph package \cite{Csardi2006}. 
	The network is then written to a file in graphml format. 
	The simulation is based on the original sourcecode of the USA/IPK model as found in the github repository\footnote{\url{https://github.com/samthiriot/model-wom-USA-IPK}}. Its execution is driven in Netlogo \cite{tisue2004netlogo} v6. 
	The OpenMole software drive the exploration of the space of parameters, the transmission of parameters and data between the generation of network and the simulation, and drives the computation in parallel. Results were analyzed in R. 
	The source code required to reproduce this exploration is shared in the same github repository.

\begin{figure}[bth]
	\includegraphics[width=\textwidth,trim={0 19pt 0 0},clip]{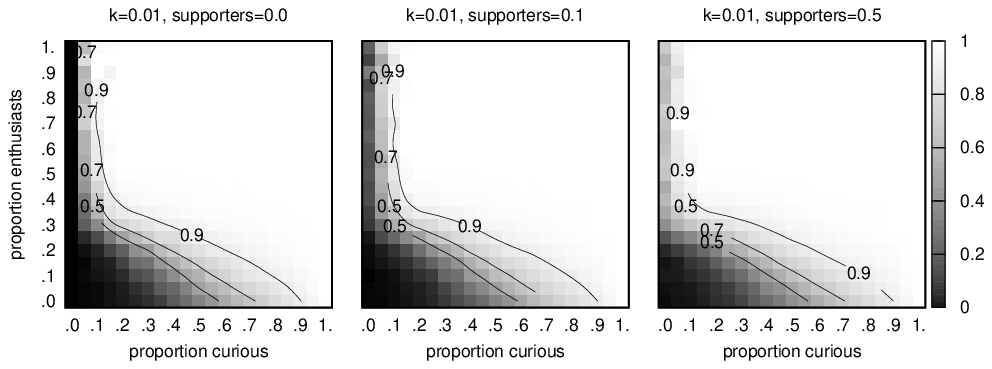}
	\includegraphics[width=\textwidth,trim={0 19pt 0 13pt},clip]{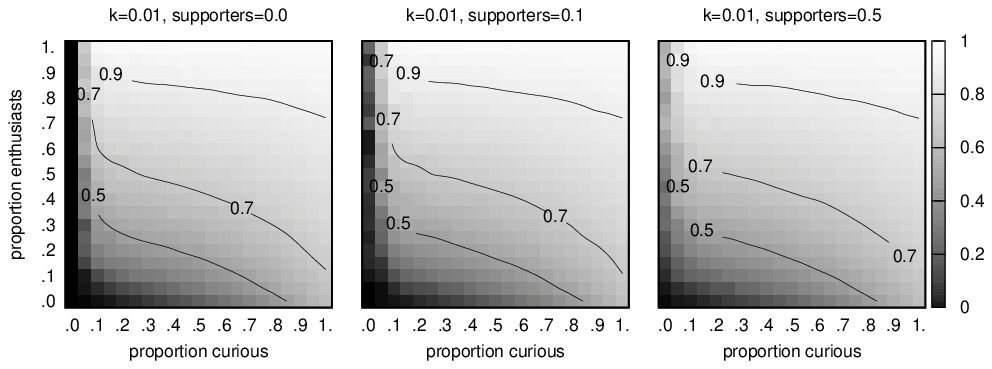}
	\includegraphics[width=\textwidth,trim={0 0 0 13pt},clip]{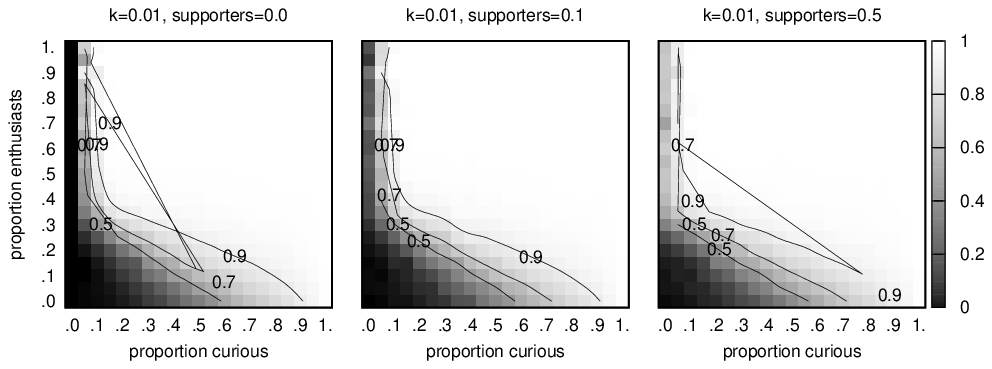}
	\vspace*{-2em}
	\caption{Final proportion of the population holding both awareness and expertise when 1\% of the population initially holds the expertise.
		\textit{(top)} WS \textit{(middle)} FF \textit{(bottom)} SII}\label{fig_heatmaps_x4_ws_k_0_01}
\end{figure} 
		
\section{Experimental results\label{xp_results}}
	
\begin{figure}[bth]
	\includegraphics[width=\textwidth,trim={0 19pt 0 0},clip]{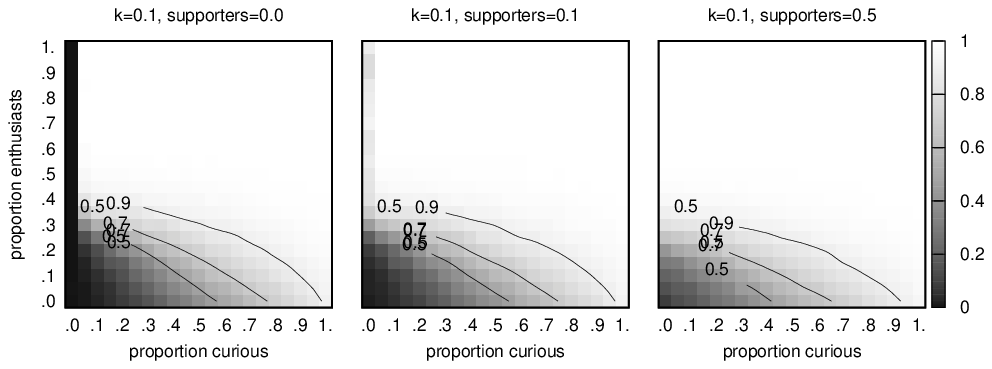}
	\includegraphics[width=\textwidth,trim={0 19pt 0 13pt},clip]{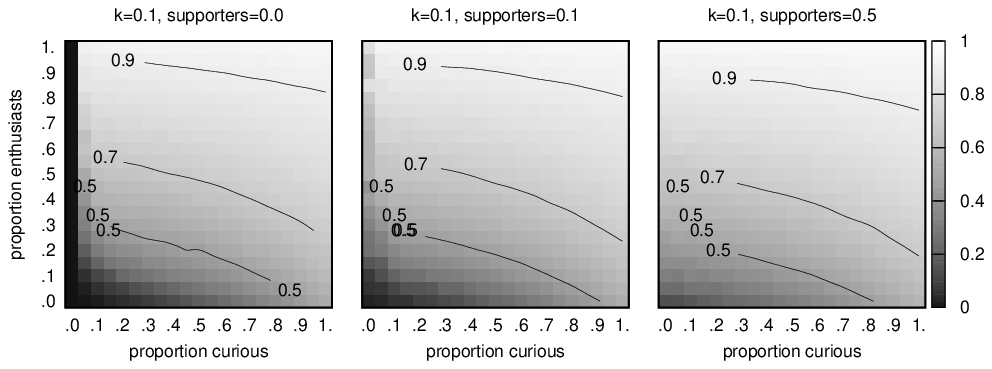}
	\includegraphics[width=\textwidth,trim={0 0 0 13pt},clip]{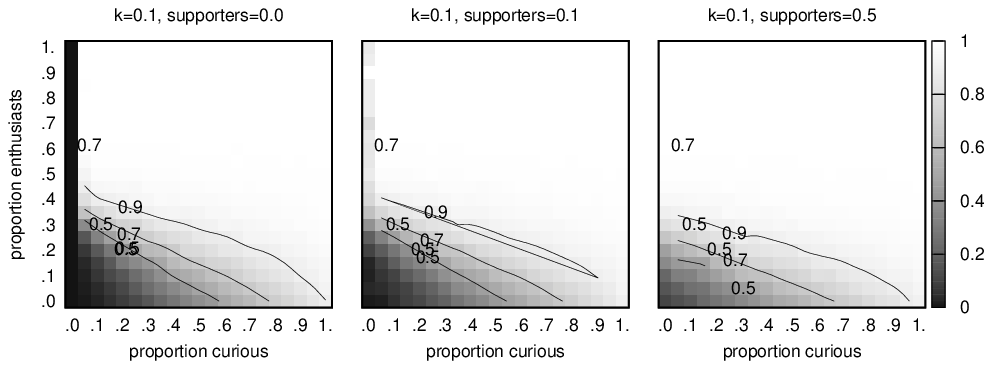}
		\vspace*{-2em}
	\caption{Final proportion of the population holding both awareness and expertise when 10\% of the population initially holds the expertise.
		\textit{(top)} WS \textit{(middle)} FF \textit{(bottom)} SII}\label{fig_heatmaps_x4_ws_k_0_1}
\end{figure}

\begin{figure}[bth]
	\includegraphics[width=\textwidth,trim={0 19pt 0 0},clip]{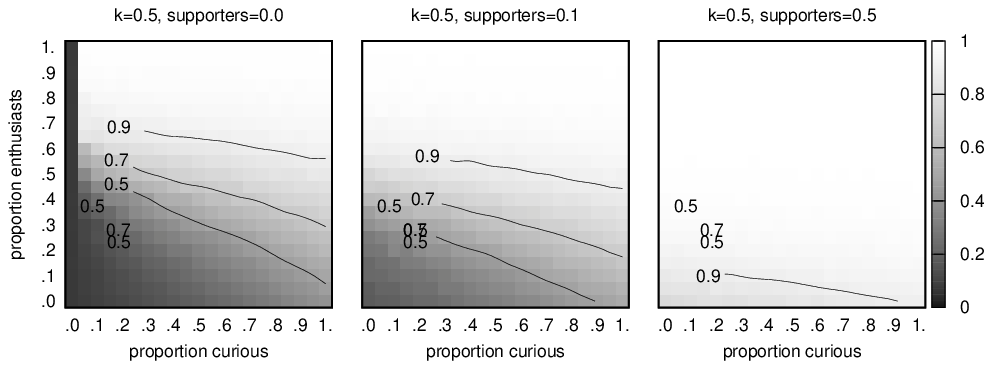}
	\includegraphics[width=\textwidth,trim={0 19pt 0 13pt},clip]{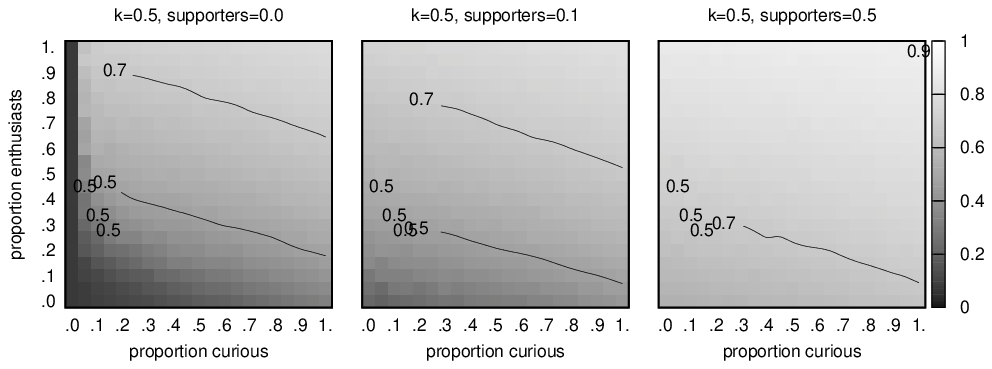}
	\includegraphics[width=\textwidth,trim={0 0 0 13pt},clip]{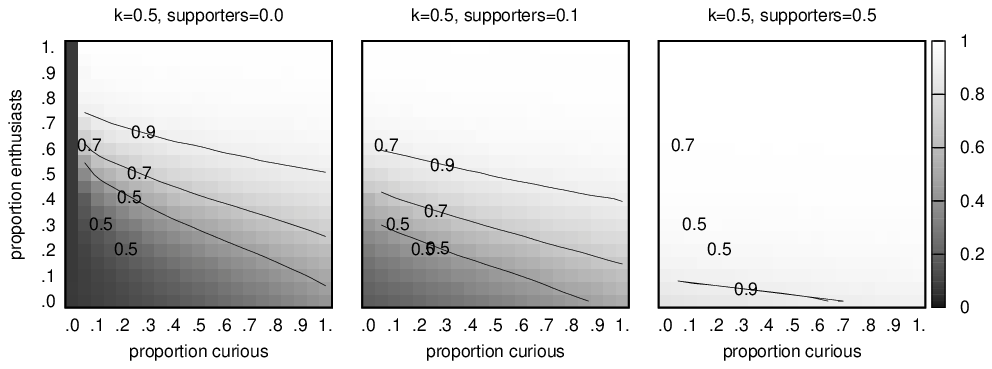}
		\vspace*{-2em}
	\caption{Final proportion of the population holding both awareness and expertise when 50\% of the population initially holds the expertise.
		\textit{(top)} WS \textit{(middle)} FF \textit{(bottom)} SII}\label{fig_heatmaps_x4_ws_k_0_5}
	\vspace{-1em}
\end{figure}

Figures~\ref{fig_heatmaps_x4_ws_k_0_01},\ref{fig_heatmaps_x4_ws_k_0_1} and \ref{fig_heatmaps_x4_ws_k_0_5} depict the simulation results for initial proportions of expertise of respectively 1\%, 10\% and 50\%. These figures represent the results of $276676$ simulations.

In the original experiment, authors noted the \textbf{high efficiency of word-of-mouth to gather the expert knowledge scattered in the population}.
The same result is obtained here on WS or SII: when there is 1\% of expertise in the population (Fig.~\ref{fig_heatmaps_x4_ws_k_0_01}), having 30\% of curious and 30\% of enthusiasts leads to more than 90\% of success. On FF however, the efficiency is noticeably lower: diffusion to the entire population is very rare. We can reproduce the original results on WS, but we also identified a counterexample with the FF networks. We conclude this efficiency depends on the structure of the network of interactions and can not be told to be systematic.

Authors highlighted \textbf{threshold effects} in the original study: a small increase in the proportion of curious or enthusiasts leads to a significant shift in the proportion of informed people. These effects are also visible here. On every network when $k=0.01$ (Fig.~\ref{fig_heatmaps_x4_ws_k_0_01}), a small increase of the proportion of curious between 0.0 to 0.05 leads to diffusion rates as different as 0 or 100\%.

Original results highlighted \textbf{three distinct regimes} of the model depending on the proportion of initial expertise $k$, with information seeking leading the dynamics when $k$ is low, symmetric roles of information seeking and supporters with $k$ higher, and first importance of supporters when $k$ is very high ($k>=0.5$). 
Thee distinct regimes also appear in our experiments depending on the value of $k$.
For $k=0.01$ (Fig.~\ref{fig_heatmaps_x4_ws_k_0_01}), there are vertical dark areas of the left side of the figures which traduce an absence of diffusion when there are no curious individuals in the population. In this part of the space of parameters, information seeking thus stands as a mandatory step for the communication to start. 
For $k=0.1$ (Fig.~\ref{fig_heatmaps_x4_ws_k_0_1}), the same effect is visible when there are no supporters (left figures), less visible with a few supporters, and not visible when there are many supporters (right figures). In this regime, curious and enthusiast agents play a similar role: at least one of them is required for the diffusion to start. 
For $k=0.5$ (Fig.~\ref{fig_heatmaps_x4_ws_k_0_5}), the success of diffusion is less sensitive to the proportion of curious agents; the supporter parameter here stands as a key element along with the proportion of enthusiasts; it would mean that when expert knowledge is widely available in the population, information seeking is less important than proactive communication. 
We confirm with these results the existence of three regimes depending on the proportion of initial expertise $k$ on WS and SII networks.

The original study points out the \textbf{strong asymmetry} in the impact of the proportions of curious and proactive agents: while diffusion can occur with only a few curious and no enthusiastic, it is not possible with only a few enthusiastic and no curious. This effect has a potential strong impact for policies, as it suggests it is more important to create campaigns of information that make people seek out for information than make people spread the word when they are knowledgeable. 
We also observe this result here: a minimal proportion of curious being required on any network and any proportion of initial expertise when there are no supporters. No other parameters plays this role in the dynamics. This suggests, in line with field observations, that information seeking is a mandatory step for diffusion of information.

In the original study, the \textbf{impact of supporters} (that is individuals who spread the word when they discover awareness \textit{after} holding the expert knowledge) only have an important impact when the initial amount of knowledge is high. We observe the same impact here: the proportion of supporters is essential for $k=0.5$, visible for $k=0.1$, and barely noticeable for $k=0.01$. This observation stands on all the networks.

The original study also highlighted how it is surprisingly \textbf{easier to retrieve expertise when there is fewer initial expert knowledge in the population}.
This finding was counter-intuitive: if more expert knowledge is disseminated in the population, then any agent seeking out information is likely to find it; the more initial expert knowledge, the more efficient the diffusion result should be. 
We find here the same results as in the initial study: with $k=0.5$, the final diffusion appears lower than with $k=0.1$ on every network. As in the initial study, this observation is explained by the diffusion process: if there are few individuals holding expertise, the people initially seeking out will create long chains of information seeking; when one agent of this chain gathers expertise, they will all collect it back (denoted "information gathering chain" in the initial study). If most people are initially knowledgeable however, as soon as an agent seeks out for information, he will find out expertise, and stop seeking out; he will not have raised the attention of many others during the process. Our experiments observe this phenomenon on every network. 

The \textbf{comparison of the dynamics of the model over different networks} mostly underlines the overall difference in the efficiency of the diffusion over networks, with diffusion being more successful over SII and WS, and far less successful over FF. As these networks have similar density and the same order of magnitude of average path length, these characteristics are unlikely to explain those differences. Other statistical indicators of the dynamics of the model also do not explain why the diffusion would be more difficult over FF networks. The duration of the diffusion is longer over WS than FF and SII, but those last two share the same distribution of duration. Observations of the dynamics of diffusion over the network suggests the core-periphery structure of FF might explain this phenomenon: the initial expert knowledge scattered in the periphery is more difficult to retrieve from the network (less dense areas). Moreover, despite the FF networks having an average path length being slightly lower than the ones of SII and WS networks in our settings, their diameter remains considerably higher, with an average of 14 instead of 7 for WS and SSI. This longer diameters makes the information less accessible in the network. 

\section{Discussion\label{discussion}}

	The experimental protocol deployed for these computational studies enabled us to successfully run hundreds of thousands of simulation, each being run over a different network whom statistical properties were measured. The use of scientific workflows to run these computations appears as a relevant scheme to analyze the impact of networks on simulated dynamics ; the combination of the OpenMole software to coordinate the exploration of the parameters, and of R/igraph to generate and analyze the networks, stands as an efficient and reliable solution.

	Regarding the USA/IPK model under study, our simulation experiments confirmed the qualitative dynamics of the model are similar over different networks: 
	importance of information seeking to trigger diffusion of innovation, asymetry between information search and proactive transmission, difficult diffusion when too much expertise is available. Any recommendation for policies based on the initial model would still stand after our computational study.
	Note that these experiments can not prove that the dynamics would be the same on any network; they just increase the likelihood of this statement. 
	However, our results demonstrated the efficiency of diffusion can be significantly different over different natures of networks (as demonstrated by the results over FF networks compared to WS and SII). 
	As for any other multi-agent model using a network to describe the structure of interactions, this raises the question of which network should be used to investigate real-world dynamics. 
	
	
	\bibliography{../../commons/biblio_bibtex/library}

\begin{thebibliography}{10}

\bibitem{samuel_thiriot:bib_customer_value:anderson_1998_1}
Eugene~W Anderson.
\newblock {Customer Satisfaction and Word of Mouth}.
\newblock {\em Journal of Service Research}, 1(1):5--17, 1998.

\bibitem{samuel_thiriot:bib_customer_value:arndt_1967_1}
J~Arndt.
\newblock {Role of product-related conversations in the diffusion of a new
  product}.
\newblock {\em Journal of Marketing Research}, 4:291--295, 1967.

\bibitem{samuel_thiriot:bib_sma_simulation:bailey_1957_1}
Norman T~J Bailey.
\newblock {\em {The Mathematical Theory of Epidemics}}.
\newblock London: Griffin, 1st edition, 1957.

\bibitem{Chatterjee1990}
R.~a. Chatterjee and J.~Eliashberg.
\newblock {The Innovation Diffusion Process in a Heterogeneous Population: A
  Micromodeling Approach}.
\newblock {\em Management Science}, 36(9):1057--1079, sep 1990.

\bibitem{Csardi2006}
G~Cs{\'{a}}rdi and T~Nepusz.
\newblock {The igraph software package for complex network research}.
\newblock {\em InterJournal Complex Systems}, 1695:1--9, 2006.

\bibitem{samuel_thiriot:bib_sma_simulation:daley_1964_1}
D~J Daley and D~G Kendall.
\newblock {Epidemics and Rumours}.
\newblock {\em Nature}, 204(4963):1118, 1964.

\bibitem{Gilly1998}
Mary~C. Gilly, John~L. Graham, Mary~Finley Wolfinbarger, and Laura~J. Yale.
\newblock {A Dyadic Study of Interpersonal Information Search}.
\newblock {\em Journal of the Academy of Marketing Science}, 26(2):83--100,
  1998.

\bibitem{samuel_thiriot:bib_sma_simulation:girvan_2002_1}
M~Girvan and M~E~J Newman.
\newblock {Community structure in social and biological networks}.
\newblock {\em Proceedings of the National Academy of Sciences}, 99(12):7821,
  2002.

\bibitem{Goffman1964}
W~Goffman and V~A Newill.
\newblock {Generalization of epidemic theory: an application to the
  transmission of ideas}.
\newblock {\em Nature}, 204:225--228, 1964.

\bibitem{samuel_thiriot:bib_sma_simulation:goldenberg_2001_1}
J~Goldenberg, B~Libai, and E~Muller.
\newblock {Talk of the Network: A Complex Systems Look at the Underlying
  Process of Word-of-Mouth}.
\newblock {\em Marketing Letters}, 12(3):211--223, 2001.

\bibitem{samuel_thiriot:bib_customer_value:goldenberg_2000_1}
J~Goldenberg, B~Libai, S~Solomon, N~Jan, and D~Stauffer.
\newblock {Marketing percolation}.
\newblock {\em Physica A: Statistical Mechanics and its Applications},
  284(1-4):335--347, 2000.

\bibitem{Granovetter1978}
M~S Granovetter.
\newblock {Threshold models of collective behavior}.
\newblock {\em American Journal of Sociology}, 83(6):1420--1443, 1978.

\bibitem{Katz1955}
Elihu Katz and Paul Lazarsfeld.
\newblock {Personal Influence: The Part Played by People in the Flow of Mass
  Communication}.
\newblock Technical report, Bureau of Applied Social Research, Columbia
  University, 1955.

\bibitem{Kempe2003}
David Kempe, Jon Kleinberg, and {\'{E}}va Tardos.
\newblock {Maximizing the spread of influence through a social network}.
\newblock {\em Proceedings of the ninth ACM SIGKDD international conference on
  Knowledge discovery and data mining - KDD '03}, page 137, 2003.

\bibitem{samuel_thiriot:bib_sma_simulation:kermack_1927_1}
W~Kermack and A~G McKendrick.
\newblock {A contribution to the mathematical theory of epidemics}.
\newblock In {\em Proc. R. Soc. Lond. A}, volume 115, pages 700--721, 1927.

\bibitem{Kiesling2012}
Elmar Kiesling, Markus G{\"{u}}nther, Christian Stummer, and Lea~M.
  Wakolbinger.
\newblock {Agent-based simulation of innovation diffusion: A review}.
\newblock {\em Central European Journal of Operations Research},
  20(2):183--230, 2012.

\bibitem{samuel_thiriot:bib_customer_value:leskovec_2007_1}
Jure Leskovec, Lada~A Adamic, and Bernardo~A Huberman.
\newblock {The dynamics of viral marketing}.
\newblock {\em ACM Trans. Web}, 1(1):5, 2007.

\bibitem{samuel_thiriot:bib_sma_simulation:leskovec_2007_1}
Jure Leskovec, Jon Kleinberg, and Christos Faloutsos.
\newblock {Graph Evolution: Densification and Shrinking Diameters}.
\newblock {\em ACM Transactions on Knowledge Discovery from Data (ACM TKDD)},
  1, 2007.

\bibitem{samuel_thiriot:bib_customer_value:meade_2006_1}
N~Meade and T~Islam.
\newblock {Modelling and forecasting the diffusion of innovation--A 25-year
  review}.
\newblock {\em International Journal of Forecasting}, 22(3):519--545, 2006.

\bibitem{Peres2010}
Renana Peres, Eitan Muller, and Vijay Mahajan.
\newblock {Innovation diffusion and new product growth models: A critical
  review and research directions}.
\newblock {\em International Journal of Research in Marketing}, 27(2):91--106,
  2010.

\bibitem{reuillon2013openmole}
Romain Reuillon, Mathieu Leclaire, and Sebastien Rey-Coyrehourcq.
\newblock {OpenMOLE, a workflow engine specifically tailored for the
  distributed exploration of simulation models}.
\newblock {\em Future Generation Computer Systems}, 29(8):1981--1990, 2013.

\bibitem{samuel_thiriot:bib_customer_value:rogers_2003_1}
Everett~M Rogers.
\newblock {\em {Diffusion of Innovations}}.
\newblock New York: Free Press, 5th edition, 2003.

\bibitem{Sheth1971}
Jagdish~N Sheth.
\newblock {Word of Mouth in Low Risk Innovations}.
\newblock {\em Journal of Advertising Research}, 11(3):15----18, 1971.

\bibitem{Thiriot2018}
Samuel Thiriot.
\newblock {Word-of-mouth dynamics with information seeking : Information is not
  ( only ) epidemics}.
\newblock {\em Physica A}, 492:418--430, 2018.

\bibitem{tisue2004netlogo}
Seth Tisue and Uri Wilensky.
\newblock {Netlogo: A simple environment for modeling complexity}.
\newblock In {\em International conference on complex systems}, volume~21,
  pages 16--21. Boston, MA, 2004.

\bibitem{Trim2012}
Kristina Trim, Naushin Nagji, Laurie Elit, and Katherine Roy.
\newblock {Parental Knowledge, Attitudes, and Behaviours towards Human
  Papillomavirus Vaccination for Their Children: A Systematic Review from 2001
  to 2011.}
\newblock {\em Obstetrics and gynecology international}, 2012:921236, 2012.

\bibitem{samuel_thiriot:bib_sma_simulation:watts_1998_1}
D~J Watts and S~H Strogatz.
\newblock {Collective dynamics of 'small-world' networks}.
\newblock {\em Nature}, 393(6684):440--442, 1998.

\bibitem{Young2009}
H.~Peyton Young.
\newblock {Innovation diffusion in heterogeneous populations: Contagion, social
  influence, and social learning}.
\newblock {\em The American Economic Review}, 99(5):1899--1924, 2009.

\end{thebibliography}

\end{document}